\documentclass[12pt,preprint]{emulateapj}

\usepackage{graphicx,amsmath,natbib}

\slugcomment{Accepted to ApJ, February 3rd, 2010}

\shorttitle{A Faint Companion to the Star $\zeta$ Vir}
\shortauthors{Hinkley et al.}

\begin{document}

\title{Discovery and Characterization of a faint stellar companion to the \\ A3V star $\zeta$ Virginis}

\author{Sasha Hinkley\altaffilmark{1,2}}
\author{Ben R. Oppenheimer\altaffilmark{3}} 
\author{Douglas Brenner\altaffilmark{3}}
\author{Neil Zimmerman\altaffilmark{3,4}}
\author{Lewis C Roberts Jr.\altaffilmark{5}}
\author{Ian R. Parry\altaffilmark{6}}
\author{R\'emi Soummer\altaffilmark{7}}  
\author{Anand Sivaramakrishnan\altaffilmark{3,8,9}}
\author{Michal Simon\altaffilmark{9}}
\author{Marshall D. Perrin\altaffilmark{10}}
\author{David L. King\altaffilmark{6}}
\author{James P. Lloyd\altaffilmark{11}}
\author{Antonin Bouchez\altaffilmark{12}}
\author{Jennifer E. Roberts\altaffilmark{5}}
\author{Richard Dekany\altaffilmark{12}}
\author{Charles Beichman\altaffilmark{13}}
\author{Lynne Hillenbrand\altaffilmark{1}}
\author{Rick Burruss\altaffilmark{5}}
\author{Michael Shao\altaffilmark{5}}
\author{Gautam Vasisht\altaffilmark{5}}

\altaffiltext{1}{Department of Astronomy, California Institute of Technology, 1200 E. California Blvd, MC 249-17, Pasadena, CA 91125}
\altaffiltext{2}{Sagan Fellow}
\altaffiltext{3}{Astrophysics Department, American Museum of Natural History, Central Park West at 79th Street, New York, NY 10024}
\altaffiltext{4}{Department of Astronomy, Columbia University, 550 West 120th Street, New York, NY  10027}
\altaffiltext{5}{Jet Propulsion Laboratory, California Institute of Technology, 4800 Oak Grove Dr., Pasadena CA 91109}
\altaffiltext{6} {Institute of Astronomy, University of Cambridge, Madingley Road, Cambridge CB3 0HA, UK}
\altaffiltext{7}{Space Telescope Science Institute, 3700 San Martin Drive, Baltimore, MD 21218}
\altaffiltext{8}{NSF Center for Adaptive Optics.}
\altaffiltext{9}{Stony Brook University}
\altaffiltext{10}{NSF Postdoctoral Fellow, UCLA Department of Astronomy}
\altaffiltext{11}{Department of Astronomy, Cornell University, Ithaca, NY 14853}
\altaffiltext{12}{Caltech Optical Observatories, California Institute of Technology, Pasadena, CA 91125}
\altaffiltext{13}{NASA Exoplanet Science Institute, California Institute of Technology, Pasadena, CA 91125}


\begin{abstract}
Through the combination of high-order Adaptive Optics and coronagraphy, we report the discovery of a faint stellar companion to the A3V star $\zeta$ Virginis. This companion is $\sim$7 magnitudes fainter than its host star in the $H$-band, and infrared imaging spanning 4.75 years over five epochs indicates this companion has common proper motion with its host star.  Using evolutionary models, we estimate its mass to be $0.168^{+.012}_{-.016}$ M$_{\odot}$, giving a mass ratio for this system $q = 0.082^{+.007}_{-.008}$. Assuming the two objects are coeval, this mass suggests a M4V-M7V spectral type for the companion, which is confirmed through integral field spectroscopic measurements.  We see clear evidence for orbital motion from this companion and are able to constrain the semi-major axis to be $\gtrsim 24.9$ AU, the period $\gtrsim 124$ yrs, and eccentricity $\gtrsim 0.16$. 
Multiplicity studies of higher mass stars are relatively rare, and binary companions such as this one at the extreme low end of the mass ratio distribution are useful additions to surveys incomplete at such a low mass ratio.  Moreover, the frequency of binary companions can help to discriminate between binary formation scenarios that predict an abundance of low-mass companions forming from the early fragmentation of a massive circumstellar disk. A system such as this may provide insight into the anomalous X-ray emission from A stars, hypothesized to be from unseen late-type stellar companions. Indeed, we calculate that the presence of this M-dwarf companion easily accounts for the X-ray emission from this star detected by ROSAT. 
\end{abstract}




\section{Introduction}
Stars with spectral type A have long shown evidence for surpising circumstellar disk structures \citep{st84,kgc05,gak06,obh08} and stars with spectral type $\sim$A6 and earlier have become increasingly targeted for low-mass companions through high-contrast imaging \citep{hhs06,hos07,hhk08,oh09} resulting in detections of several low-mass companions \citep{kgc08,mmb08}.  Indeed, \citet{jbm07} suggest that the frequency of {\it planet} occurrence around A-type stars is twice that of solar-mass stars.  In addition to planetary-mass companions, the frequency and mass ratio distributions of {\it stellar}-mass companions to nearby A stars can help constrain binary formation scenarios---such as models based on the more massive primary star dynamically capturing a lower mass companion \citep{mc93}, or a picture relying on initial fragmentation within a protostellar cloud, e.g. \citet{bb96}.  
Although some multiplicity studies of A and B stars have been conducted---e.g. the \citet{st02} and \citet{kbz05} surveys of Sco OB2---a comprehensive statistical picture of multiplicity around these massive stars, based on both cluster and field objects, has yet to emerge.   Observations of massive, early-type stars may serve as important boundary-type systems, to which models of formation must conform. Specifically, an abundance of brown dwarf/M-dwarf companions to A stars would lend support to recent models describing the formation of these objects through the fragmentation of an initially massive circumstellar disk \citep{kmy09,sw09}.  

Moreover, the frequency of stellar companions to A-stars may be related to their anomolous source of X-rays.  Since A-stars have shallow or non-existent convective regions in their envelopes, they lack a significant dynamo effect, and can be expected to display negligible X-ray emission. Meanwhile, M dwarfs are well known sources of X-rays \citep{fgs93}.  
\citet{sgh85} suggested that unseen late-type stellar companions to A stars may be the source of the X-rays. Indeed, \citet{ss07} find that the majority of nearby X-ray emitting A-stars have some evidence of possessing low-mass companions, likely responsible for the X-ray emission. Moreover, \citet{zoh10} have recently discovered a mid-M dwarf companion to the nearby A-star Alcor, a ROSAT source.

\begin{deluxetable}{llc}
\tabletypesize{\scriptsize}
\tablecaption{Fundamental Parameters for $\zeta$ Virginis A}
\tablewidth{0pt}
\tablehead{\colhead{Parameter}  & \colhead{} & \colhead{Value}}
\startdata                                     
         Identifiers                                                                                               & & HIP66249, HR5107,\\
                                                                                                                          & & HD118098 \\
         Spectral Type\footnote{\citet{gcg03}}                                               & & A3V        \\ 
         $V$ magnitude\footnote{\citet{hj82}}                                                & & 3.40   \\
         Parallax (mas)\footnote{\citet{plk97}}                                               & &  44.55$\pm0.90$\\
         RA, Dec Proper Motion (mas yr$^{-1}$)\footnote{\citet{plk97}}   & & -278.89$\pm0.83$, \\
                                                                                                                          & &  48.56$\pm0.71$ \\
         Radial Velocity (km s$^{-1}$)\footnote{\citet{ab72, ldg09}}          & & -13.2$\pm0.3$ \\
\enddata
\label{fundpar}
\end{deluxetable}

\subsection{Previous Studies of $\zeta$ Virginis}
The star $\zeta$ (``Zeta'') Virginis (HIP66249, A3V, $V$=3.40---See Table~\ref{fundpar}, hereafter $\zeta$ Vir), is a target in our on-going high-contrast imaging program \citep{odn04,soh07,hob08}.  This nearby (22.45 pc) star has been previously used as a calibrator star for interferometric work \citep{acm09}, and has also been followed with the HARPS survey for radial velocity variations \citep{ldg09}. No such variations were found. The Bright Star Catalogue \citep{hj82} lists $\zeta$ Vir as a member of the Hyades moving group. There is some spread in the derived age of this group. Although some authors claim ages as high as 625 Myr \citep{pbl98} or 650 Myr \citep{lfl01}, \citet{rss05} cites the age of this star at 505 Myr. Rather than trying to establish the membership of this star in the Hyades moving group, and then adopt the moving group age for the star, we simply adopt the 505 Myr age of \citet{rss05} for the analysis in this paper. 
 $\zeta$ Vir has indeed been observed by ROSAT \citep{hsv98}, and is listed as a single star, with an X-ray brightness ($L_x=1.07\times 10^{28}$ erg s$^{-1}$). Despite the fact that there is a 20$^{\prime\prime}$ offset between X-ray and optical positions, the ROSAT catalog lists a 14$^{\prime\prime}$ uncertainty on the position of $\zeta$ Vir. Such a 1.5$\sigma$ positional offset easily leaves open the possibility that the observed X-ray source is in fact located at $\zeta$ Vir. 
The $\zeta$ Vir system has a radial velocity of -13.2 km s$^{-1}$ as mentioned in \citet{ab72} and \citet{kok98} found this radial velocity to be constant at the 1-2 km/s level, i.e. lacking a companion, using it as one of their standard calibrator stars.

Speckle observations at the Canada-France-Hawaii Telescope in the optical \citep{mmh93} did not find a binary companion for this star, however this is not surprising given the comparatively low dynamic range ($\Delta M$$\sim$3 mags) of this technique. \citet{pmm01} specifically carried out a survey of bright early-type stars both in the field and in clusters to search for companions using AO, but no mention of this star is listed.

\section{Observations}
We have imaged the $\zeta$ Vir system using two observing programs, each with different instruments: A coronagraph working together with an infrared camera, and a newly commisioned coronagraph which employs an IFS as the primary science camera. We describe each observing program below.

\subsection{``The Lyot Project'', a Coronagraphic Imager at AEOS}
The first instrument, ``The Lyot Project'' \citep{odn04,soh07} was a diffraction-limited classical Lyot coronagraph \citep{l39,skm01} working with the Adaptive Optics (hereafter ``AO'') system on the 3.63 m AEOS telescope on Haleakala, Hawaii \citep{rn02}.  Our images were gathered using ``Kermit,'' an infrared camera \citep{mdp03}, with a 13.5 mas pixel$^{-1}$ plate scale, and differential polarimetry mode for detection of diffuse circumstellar material \citep{hos09}. Images of $\zeta$ Vir in $J$, $H$, and $K$-bands were obtained using this instrument over three epochs spanning three years as listed in Table~\ref{observations}.  
Our coronagraph used a focal plane mask with a 455 mas diameter (4.9$\lambda/D$ at $H$-band), as well as its own internal tip/tilt system.  Images in the $J$, $H$, and $K$-bands were obtained during the second and third epochs, while only $H$-band was obtained on the first. To calibrate our photometry, we also obtained 1 s unocculted images in addition to the coronagraphically occulted images.  In this setup, the target is more than 1$^{\prime\prime}$ away from our occulting mask.  
The raw data images, both occulted and unocculted, required a mix of both traditional data reduction steps (e.g. dark current subtraction, bad-pixel masking, flat-field correction) as well as some techniques customized for the infrared camera.  More details on the data reduction are given in \citet{soh06} and  \citet{dho06}.


\begin{deluxetable*}{lccccc}
\tabletypesize{\scriptsize}
\tablecaption{Observations and Astrometric Distance Between Primary and companion for the $\zeta$ Vir System}
\tablewidth{0pt}
\tablehead{\colhead{Epoch}  & \colhead{MJD}  & \colhead{Wavelength} & \colhead{Observing Method} & \colhead{Separation (mas)} & \colhead{PA (degrees East of North)}  }
\startdata                                     
         1) & 53168.2839 & $H$-band               & Lyot Coronagraph   &1846 $\pm$ 9   & 144.7 $\pm$ 0.1        \\ 
         2) & 53507.3742 & $J$, $H$, $K$-bands    & Lyot Coronagraph   &1830 $\pm$ 3   & 146   $\pm$ 1.0        \\ 
         3) & 54257.5378 & $J$, $H$, $K$-bands    & Lyot Coronagraph   &1814 $\pm$ 16  & 147.4 $\pm$ 0.1       \\ 
         4) & 54657.1548 & 1.1 - 1.8 $\mu$m (IFS) & APLC + IFS         &1790 $\pm$ 12  & 149.8 $\pm$ 0.1   \\ 
         5) & 54904.4445 & 1.1 - 1.8 $\mu$m (IFS) & APLC + IFS         &1779 $\pm$ 12  & 151.0 $\pm$ 0.2    \\ 
\enddata
\label{observations}
\end{deluxetable*}

\subsection{``Project 1640,'' a coronagraphic Integral Field Spectrograph at Palomar}
The second instrument used to image the $\zeta$ Vir system is a new instrument \citep{hob08} recently commissioned on the 200-in Hale Telescope at Palomar Observatory. This instrument, termed ``Project 1640,'' is a coronagraph integrated with an integral field spectrograph (IFS) spanning the $J$ and $H$-bands (1.05$\mu$m - 1.75$\mu$m). The IFS+Coronagraph package is mounted on the Palomar AO system \citep{dbp98}, which in turn is mounted at the Cassegrain focus of the Hale Telescope.  The coronagraph is an Apodized-Pupil Lyot coronagraph (APLC) \citep{s05}, an improvement of the classical Lyot coronagraph \citep{skm01}. This coronagraph uses a 370 mas diameter (5.37$\lambda/D$ at $H$-band) focal plane mask. The IFS, or hyper-spectral imager, is a microlens-based spectrograph which can simultaneously obtain $4\times10^4$ spectra across our $4^{\prime\prime}\times4^{\prime\prime}$ field of view. Each microlens subtends 19.2 mas on the sky and a dispersing prism provides a spectral resolution ($\lambda/\Delta\lambda$)$\sim$32. 

The IFS focal plane consists of $4\times10^4$ spectra that are formed by dispersing the pupil images created by each microlens.  To build a data cube, the data pipeline uses a library of images made in the laboratory with a tunable laser, which spans the operating wavelength band of the instrument.  Each image contains the response of the IFS to laser emission at a specific wavelength---a matrix of point spread functions.  Each laser reference image is effectively a key showing what regions of the $4\times10^4$ spectra landing on the detector correspond to a given central wavelength.  Each laser reference image is cross correlated with the focal plane spectra to extract each wavelength channel.   

The pixel scales for each instrument were calculated at by imaging four binary stars (HIP107354, HIP171, HIP88745, and WDS11182+3132) with high quality orbits with small astrometric residuals \citep{hmw01}. 
The pixel scale is calculated by performing a least squares fit between these predicted separations and the pixel separation in our data.  

\section{The Companion}
\begin{figure}[ht]
\center
\resizebox{1.2\hsize}{!}{\includegraphics{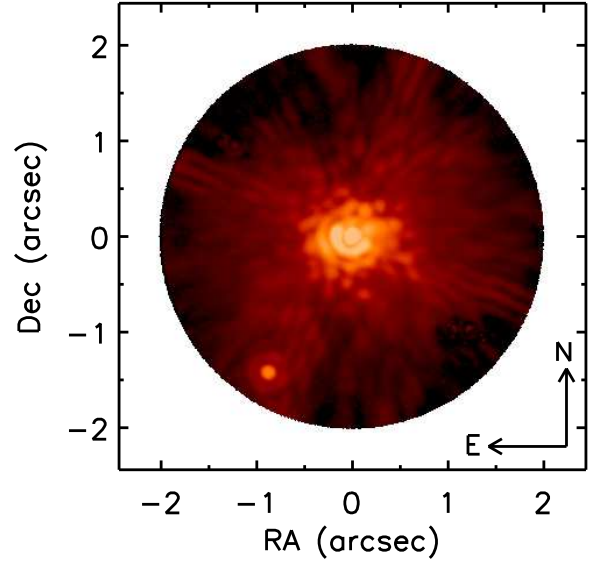}}
  \caption{A 60 s $H$-band (1.65 $\mu$m) image of the the star $\zeta$ Vir taken on 2004 June 12 (UT) at the AEOS telescope. In this discovery image, the adaptive optics system is on, and the star has been coronagraphically occulted behind our 455 mas occulting mask.  A faint stellar companion,  $\zeta$ Vir B, 7 magnitudes fainter than the host star and sharing common proper motion with the host star, is apparent at the bottom left of the image.}
 \label{zetavir} 
\end{figure}

Here we report the discovery of a faint stellar companion to $\zeta$ Vir, hereafter $\zeta$ Vir B. The discovery image is shown in Figure~\ref{zetavir}.  To our knowledge, the existence of this companion has not been reported previously.

\subsection{Common Proper Motion Analysis}
\begin{figure}[!ht]
\center
\resizebox{1.0\hsize}{!}{\includegraphics{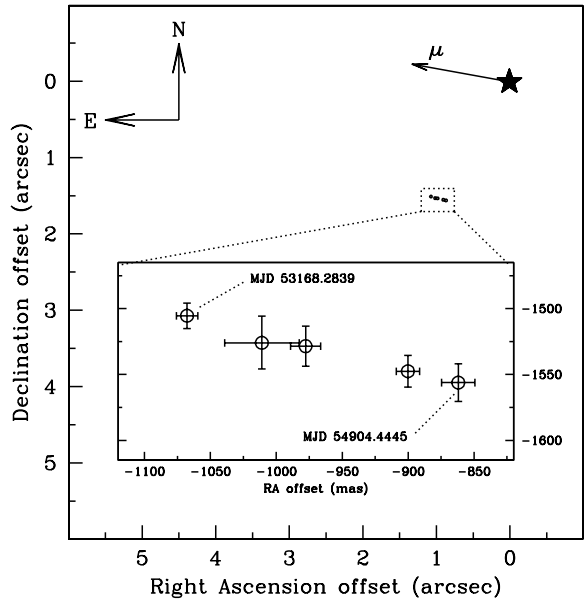}}
  \caption{The offset positions of $\zeta$ Vir B relative to the host star.  The position of the host star is marked with the $\star$ symbol.  The inset portion of the plot shows the positions of the stellar companion over the 4.75 years of observations presented in this paper.  The error bars incorporate the uncertainties in the radial separation and the position angle.  The vector labelled ``$\mu$'' at the upper right shows the magnitude  and direction of the proper motion of the $\zeta$ Vir system over the 4.75 year duration of these observations (-1325 mas/yr, 230.66 mas/yr).  }
  \label{zetavirpos} 
\end{figure}

The astrometric measurements for the primary/companion separation are presented in Table~\ref{observations}. For the first and third epochs, the astrometric positions of the two stars were obtained from images in which the primary star was occulted. In these cases, a centroid to each PSF was calculated as part of a best fit radial profile measurement.  With coronagraphic imaging, the exact position of the occulted star is difficult to determine. The uncertainty can be estimated using binary stars, in which one of the binary members is occulted \citep{dho06}.  For all but the first epoch, we used a physical mask with a superimposed grid \citep{so06}, which produces fiducial reference spots indicating the position of the host star to within $\sim$10 mas.  The second, fourth and fifth epochs contained fully unocculted data with sufficiently high signal-to-noise to measure the position of both the primary and the companion. $\zeta$ Vir A, listed as a high-proper motion star, has a proper motion of 283 mas yr$^{-1}$ \citep{plk97}. If these two objects were not associated with each other, we could expect a $\sim$$1.35^{\prime\prime}$ change in separation over the 4.75 year course of observations. Instead, we report a $\sim$200 mas southwesterly change in the position of the companion (see Figure~\ref{zetavirpos}) relative to the host star. Since the relative separation between the host star and the companion is far less than the $\sim$$1.35^{\prime\prime}$ change in separation if the two were not mutually bound, we are confident that these two objects share common proper motion.  Moreover, this westerly change reflects the orbital motion of $\zeta$ Vir B over the 4.75 year observing baseline.

\subsection{Photometry}
Aperture photometry of the companion was performed on images from the Lyot Project in which $\zeta$ Vir was occulted behind our coronagraphic mask. The flux was summed in apertures of radii of 270 mas, 340 mas, and 300 mas, for the $J$, $H$, and $K$-band images respectively, followed by sky subtraction. Zeropoints for the $J$, $H$ and $K$ data were derived by performing large (760 mas) aperture photometry on unocculted, unsaturated images of $\zeta$ Vir A taken immediately prior to the occulted observations. We calculate $J$, $H$, and $K$-band photometric zero points of $20.990 \pm 0.017$, $20.639 \pm 0.078$, and $19.932 \pm 0.069$, respectively.  Assuming a distance of 22.45 pc \citep{plk97}, Table~\ref{phot} shows our calculated absolute $J$, $H$, and $K$-band magnitudes of $8.99\pm .06$, $8.41\pm.14$, and $8.14\pm.17$, respectively for $\zeta$ Vir B.

\subsection{Spectroscopy}
Integrating an IFS into more conventional high-contrast imaging techniques can provide significantly more information on objects in close vicinity to their host star \citep{sf02,bgf06,mml07}. Normally, when spectra are unavailable, parameters such as mass, spectral type, and age must be derived by combining broadband photomery with model predictions. Such models can be problematic at very young ages \citep{smv06,ajl07}. 

\begin{figure}[ht]
\center
\resizebox{1.15\hsize}{!}{\includegraphics{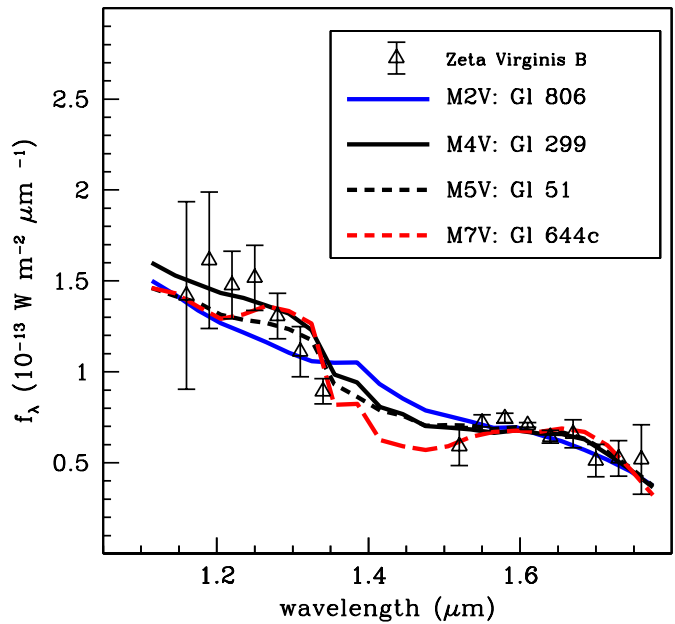}}
  \caption{The $J$ and $H$-band spectrum of $\zeta$ Vir B obtained with our IFS and coronagraph at Palomar \citep{hob08}. Also shown are template spectra for M2V through M7V stars taken from the IRTF spectral Library \citep{crv05,rcv09}. The water band data points between $\sim$1.35 and $\sim$1.5$\mu$m have been excluded due to the variation of this band between observations of $\zeta$ Vir and our calibrator star. }
  \label{spectra}
\end{figure}

\begin{figure*}
\center
\resizebox{1.05\hsize}{!}{\includegraphics{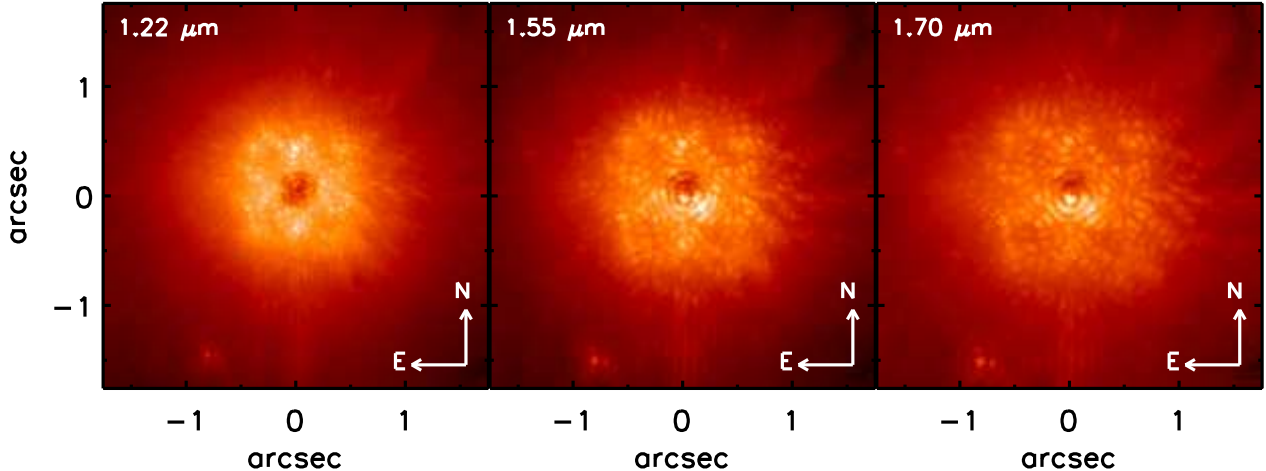}}
  \caption{Three images of the $\zeta$ Vir system taken from a data cube from the Project 1640 IFS.  The three images show the system at 1.22, 1.55, and 1.70 $\mu$m. The host star, $\zeta$ Vir A has been coronagraphically occulted at the center of each image, and the companion, $\zeta$ Vir B is evident at the lower left. }
  \label{zetavir1640}
\end{figure*}

Observations with our IFS at Palomar Observatory \citep{hob08} allowed us to obtain the spectrum of $\zeta$ Vir B shown in Figure~\ref{spectra}. Each point in the spectrum of $\zeta$ Vir B was calculated by performing aperture photometry on each image in a data cube. Examples of three such images taken from a data cube are shown in Figure~\ref{zetavir1640}.  The photometry was obtained using a circular aperture such that the second Airy ring of the Point Spread Function was enclosed at each wavelength in the data cube. A median background sky value was calculated in a 40 mas wide annulus, just outside the photometric aperture, and subtracted from the target counts.  Each flux value for $\zeta$ Vir B shown in Figure~\ref{spectra} is a median of five data points taken from five data cubes of $\zeta$ Vir B, and the error bars show the 1$\sigma$ spread of these five values.  The spectrum was calibrated using a reference star (HIP 56809, G0V, $V$=6.44) by comparing the measured counts of the reference star with a template G0V star (HD 109358, G0V, $V$=4.26) taken from the IRTF spectral Library \citep{crv05,rcv09} to derive a spectrograph response function.  

\begin{deluxetable}{ccc}
\tabletypesize{\scriptsize}
\tablecaption{Photometry for $\zeta$ Virginis B}
\tablewidth{0pt}
\tablehead{\colhead{Band}  & {apparent magnitude} & \colhead{absolute magnitude}}
\startdata                                     
         $J$       &  $10.75\pm 0.06$  & $8.99\pm 0.06$ \\
         $H$      &  $10.17\pm 0.14$ & $8.41\pm 0.14$ \\
         	$K$      &  $9.90\pm 0.17$  & $8.14\pm0.17$ 
\enddata
\label{phot}
\end{deluxetable}

The spectrum shown in Figure~\ref{spectra} has had this response correction applied to it. We have excluded the data points in the vicinity of the water absorption band between $\sim$1.35 $\mu$m and $\sim$1.5$\mu$m, since the degree of water absorption present in the calibrator star was sufficiently different from that present in the $\zeta$ Vir observations.  Also shown in the figure are template spectra for an M2V through M7V star taken from the IRTF spectral Library.  The extracted spectrum for $\zeta$ Vir B is most consistent with the M4V - M7V  spectral types.

\section{Analysis}
 \subsection{Mass and Age of $\zeta$ Vir A}
\begin{figure*}[ht]
\center
\resizebox{1.0\hsize}{!}{\includegraphics{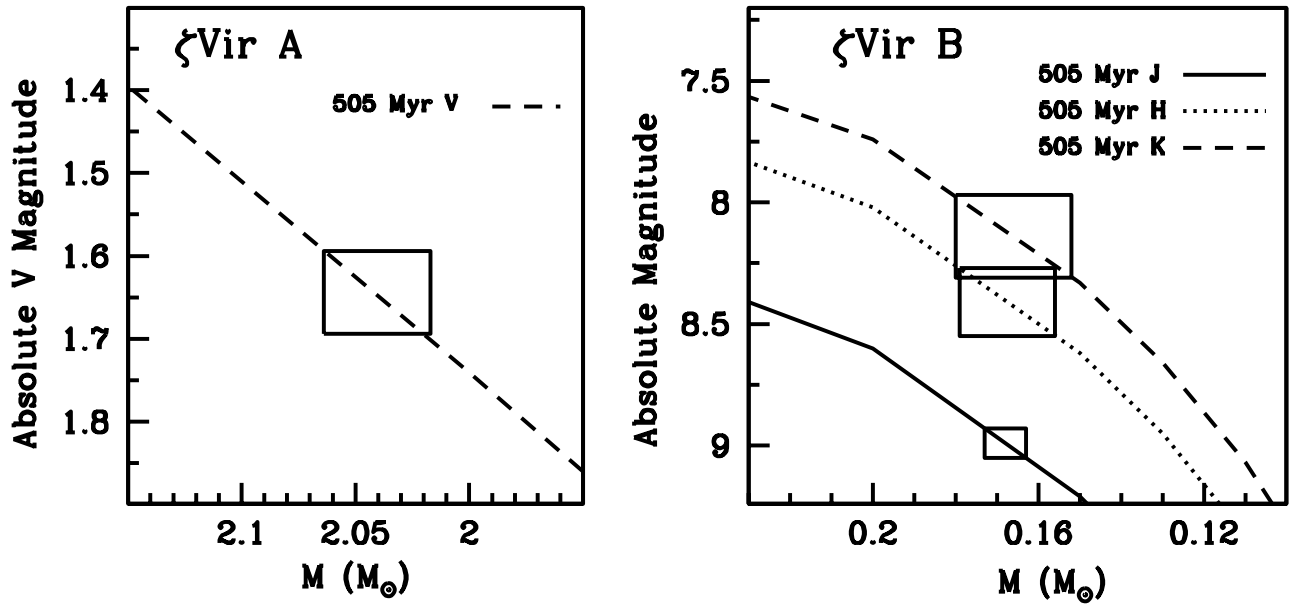}}
  \caption{The left hand panel shows the luminosity-mass relation for the host star, $\zeta$ Vir A as calculated by \citet{sdf00} for a 505 Myr system.  We use  absolute $V$-magnitude in lieu of total luminosity.  The vertical extent of the box indicates the $V$-band photometric uncertainty ($M_V = 1.64\pm 0.05$) for $\zeta$ Vir A. This defines an allowable region for the mass for the A3V host star to be $2.041\pm .024$ M$_\odot$. The right panel shows evolutionary models for low-mass stars taken from \citet{bcb03} for the $J$, $H$, and $K$-bands.   As with the case for $\zeta$ Vir A, the vertical extent of each box indicates the photometric uncertainty at each band.  These three bandpass values gives an overall value of $0.168^{+.012}_{-.016}$ M$_\odot$ for $\zeta$ Vir B.  }
  \label{magmass}
\end{figure*}

Although \citet{hj82} list $\zeta$ Vir as a possible member of the Hyades moving group, several lines of evidence suggest simply assigning the system the age of the Hyades cluster is not rational.  Indeed, using criteria based on mass distribution and metallicity, \citet{fpl07} question whether most members of the Hyades Moving Group are actually evaporated members of the Hyades Open Cluster.  The mean metallicity of the Hyades cluster has been well established with an [Fe/H] value of $0.14\pm0.05$ \citep{ccl97}, and later refined to $0.144\pm 0.013$ \citep{g00}.  However, the mean metallicity value of [Fe/H] $\simeq -0.02$ \citep{gcg03} for the $\zeta$ Vir system is significantly different than the above values, ruling out the possibility that this star is a member of the Hyades cluster.
For this work, we have decided to adopt the age of 505 Myr given by \citet{rss05}. 
In the left panel of Figure~\ref{magmass} we show the luminosity-mass parameter space showing model tracks calculated by \citet{sdf00} for a 505 Myr system with solar metallicity.  The vertical extent of the box indicates the absolute $V$-band photometric uncertainty. This uncertainty in the luminosity ($M_V = 1.64\pm 0.05$), defines an allowable mass region for $\zeta$ Vir A of $2.041\pm .024$ M$_\odot$.

\subsection{Mass and Age of $\zeta$ Vir B}

We use the models of \citet{bcb03} to derive a mass for $\zeta$ Vir B assuming the age of 505 Myr. In Figure ~\ref{magmass} (right panel) we show plots of the $J$, $H$, and $K$-band absolute magnitudes for a range of  companion masses calculated from these models.  
As with the case  of $\zeta$ Vir A, the uncertainty in the photometry of this object has very little effect on the derived mass of the companion. 
Together these values give an overall derived mass of $0.168^{+.012}_{-.016}$ M$_\odot$. We take this value as the final derived mass for $\zeta$ Vir B. 



To check the validity of this model-based mass estimate, we compare this value with empirically derived mass-luminosity relations given in \citet{hm93} and \citet{dfs00}. Using the $J$, $H$, and $K$ magnitudes, these two works predict values of $0.152 \pm 0.009$ M$_\odot$ and $0.166 \pm 0.004$ M$_\odot$ (See Table~\ref{masstable}), consistent with a mid-M spectral type for a main sequence star at these ages \citep{rhg95,hgr96}, and consistent with the spectral determination derived previously.  
The \citet{hm93} and \citet{dfs00} mass values are slightly lower than the model-based 0.168 M$_\odot$ value given by the \citet{bcb03} models, but are still consistent with a mid-M spectral type for $\zeta$ Vir B.
Using the range of primary star masses derived above gives this system a mass ratio of $q = 0.082^{+.007}_{-.008}$. 


\begin{deluxetable*}{llcccc}
\tabletypesize{\scriptsize}
\tablecaption{$J$,$H$,and $K$-band Mass Determinations for  $\zeta$ Vir B.}
\tablewidth{0pt}
\tablehead{\colhead{Model}  & \colhead{Method} & \colhead{$J$-band} & \colhead{$H$-band} & \colhead{$K$-band} & \colhead{Median$\pm$1$\sigma$}}
\startdata                                     
    \citet{hm93}                      &   Mass-Luminosity (Empirical)   &   0.136 M$_\odot$    & 0.152 M$_\odot$            & 0.152 M$_\odot$     & $0.152\pm 0.009$ M$_\odot$      \\
    \citet{bcb03} (505 Myr)  &  Evolutionary (Theoretical)     &   0.168 M$_\odot$    & 0.169 M$_\odot$            & 0.167 M$_\odot$     & $0.168\pm 0.001$ M$_\odot$        \\
    \citet{dfs00}                       & Mass-Luminosity (Empirical) &   0.170 M$_\odot$    & 0.166 M$_\odot$            & 0.163 M$_\odot$     &  $0.166\pm 0.004$ M$_\odot$         \\ 
\enddata
\label{masstable}
\end{deluxetable*}

\subsection{Orbital Analysis}

\begin{figure*}[ht]
\center
\resizebox{.9\hsize}{!}{\includegraphics{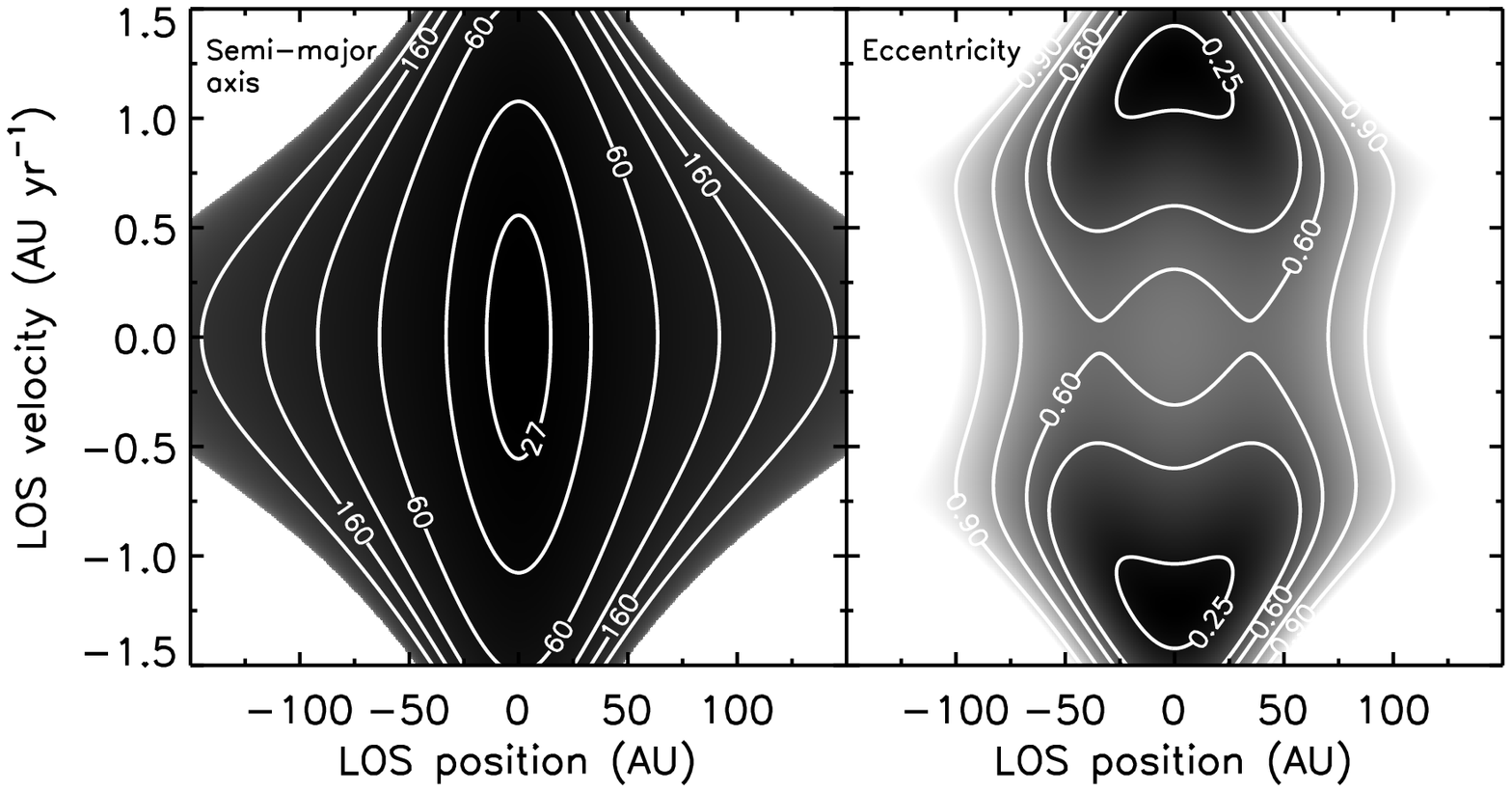}}
  \caption{The loci of possible values of the semi-major axis (left), and eccentricity (right), for $\zeta$ Vir B, assuming a range of line-of-sight positions and velocities \citep{gbk98}. The contours for the left plot are 27, 35, 60, 100, 160, and 300 AU. The right hand plot  shows eccentricity contours equal to 0.25 0.50, 0.60, 0.75, and 0.90. This constrains the semi-major axis to be $\gtrsim 24.9$ AU (and hence the period to be $\gtrsim124$ yrs), and eccentricity $\gtrsim 0.16$.}
  \label{orbconstrain}
\end{figure*}

Given the relatively short span (4.75 years) of the observations of $\zeta$ Vir B (see Figure~\ref{zetavirpos}), fitting an orbital model to the data is premature. However, we may borrow an analysis used by \citet{gbk98} to constrain the eccentricity, $e$, and semi-major axis, $a$, of $\zeta$ Vir B using our astrometry in the two-dimensional plane of the sky, combined with Kepler's Laws. We refer the reader to \citet{gbk98} for a full explanation.  Over our 4.75 year time baseline, we calculate a velocity of -1.139 AU yr$^{-1}$ in the westward direction, and .062 AU yr$^{-1}$ in the south direction.  Assuming the orbit of $\zeta$ Vir B is bounded, and assuming a range of line-of-sight positions and velocities for $\zeta$ Vir B, we are able to constrain $a$ and $e$, and we show the loci of possible values for these parameters in Figure~\ref{orbconstrain}.  The ordinate and abcissa values show the assumed values of the line-of-sight velocity, and position, respectively.  The values shown in Figure~\ref{orbconstrain} indicate $a\gtrsim 24.9$ AU and $e\gtrsim 0.16$. From the semi-major axis constraint, we can constrain the period $P=(4\pi^2a^3/\mu)^{1/2}$ to be $\gtrsim 124 $ yrs, where $\mu = G(m_1 + m_2)$.  In this analysis we have assumed a mass of $m_1=2.04 M_\odot$ (see Figure~\ref{magmass}) and $m_2 = .168 M_\odot$ for $\zeta$ Vir A and $\zeta$ Vir B, respectively.

\section{Discussion}
Assuming a mass of $2.041\pm .024$ M$_\odot$ for $\zeta$ Vir A gives this newly discovered binary system a mass ratio of $q = 0.082^{+.007}_{-.008}$.  Although numerous low-mass companions have been detected around $\sim$1 M$_\odot$ stars, this system is of particular interest given that it orbits a primary star of $\sim$2 M$_\odot$.  
The $q$ distributions for primaries in the mid and low mass stellar regimes have been well studied, but comprehensive binary statistics for A stars are incompletely surveyed.
The mass ratio distribution for the mid and lower mass ranges show fairly clear trends with mass. Namely,  \citet{brs07} discusses that 
very low mass binaries have mass ratios that are skewed towards unity.
Studies using a complete sample of stars between 0.6 - 0.85 M$_\odot$ \citep{msp03}, as well as M-dwarf surveys \citep{fm92,rg97}, find a significantly flat distribution of mass ratios.
In the same vein, \citep{kim08}, used a sample of 82 young stars of GKM type in the Upper Sco star forming region and found a distribution of mass ratios not significantly different from a constant distribution, i.e. not significantly biased towards having low mass companions.   Towards higher masses, \citet{dm91} find that F and G type stars have a broad range of mass ratios, but with a slight increase towards small secondary masses.  Finally, at the higher mass end, \citet{st02} and \citet{kbz05} have performed a survey of A and B-stars in the Sco OB2 association, finding a high rate of binarity. 

Similarly, many studies show a positive correlation between the distribution of binary separations and the mass of the system. 
The \citet{dm91} sample of solar mass stars show a mean separation of $\sim$30 AU. At lower masses, the \citet{fm92} survey (largely M dwarfs within 8 pc) and \citet{rg97} surveys (M dwarfs within 20 pc) find mean separations between 4 and 30 AU.  Continuing towards lower mases,  low mass M dwarfs and brown dwarfs as discussed in \citet{brs07}, have notably smaller mean separations ($\sim$4 AU), with maximum separations $\sim$20 AU.

\citet{kbz05} and \citet{st02} have undertaken the first steps towards a comprehensive study of the multiplicity of A stars. Our work aims to aid in that effort.  The value of the current study lies in the ability to obtain a spectrum which determines the spectral type, a tight constraint on the secondary mass, and a constrained orbit.  
 

Our finding may also be useful for studies of the anomalous X-ray emission from A stars.  Unseen late-type companions to A stars have been hypothesized to be the source of their anomolous X-ray emission.  
Indeed, the presence of the M-dwarf companion in this system can easily account for the X-ray flux detected by ROSAT. For a $.168 M_\odot$ star at 505 Myr, the \citet{bcb03} evolutionary tracks predict a luminosity of  $L_{bol} \simeq 2\times10^{31}$ erg s$^{-1}$.  And assuming that the X-ray luminosity $L_{x} = 1.07\times10^{28}$ erg s$^{-1}$ noted by \citet{hsv98} is due completely to the companion, this predicts a $\log(L_{x}/L_{bol})$ value of -3.3, quite typical for a young mid M dwarf (See \citet{fgs93}, especially their Figure 3).  Recently, \citet{zoh10} have reported the presence of a mid-M dwarf bound to the star Alcor.  More complete high-contrast surveys for companions surrounding an ensemble of A stars will allow researchers to begin to address the issue of the anomolous X-ray emission in a statistically robust manner. 


An abundance of M-dwarf companions in configurations like this also may lend support to models of binary formation based on fragmentation of a massive circumstellar disk. As \citet{kmy09} point out, massive stars with their presumably massive circumstellar disks and correspondingly high mass infall rates provide an environment conducive for the formation of disk instabilities and fragmentation. Indeed, such a mechanism is more likely for disks surrounding stars more massive than 1-2M$_\odot$\citep{kkm07,kmk08}.
If indeed this fragmentation is a prominent mechanism for the formation of binary companions \citep{sw09}, the abundance of low mass companions (brown dwarfs and M-dwarfs) should be more frequent around more massive stars.

\section{Summary}
We report the discovery of a low-mass, M4V-M7V stellar companion to the star $\zeta$ Vir.  This object clearly shares common proper motion with its host star, and we derive a mass of $0.168^{+.012}_{-.016}$ M$_{\odot}$, corresponding to a mass ratio $q = 0.082^{+.007}_{-.008}$. Our broad-band photometry and spectroscopy are consistent with an mid-M spectral type.  Although numerous low-mass companions have been identified around $\sim$1 M$_\odot$ systems, this object is significant given its membership in a  $\sim$2 M$_\odot$ system. Characterization of more systems like this are important for identifying the anomalous source of X-rays from A stars as well as constraining possible modes of formation of stellar companions through the fragmentation of massive circumstellar disks.

\acknowledgments
We thank the anonymous referee for his or her comments. 
This work was performed in part under contract with the California Institute of 
Technology (Caltech) funded by NASA through the Sagan Fellowship Program.
The Lyot Project is based upon work supported by the National Science  
Foundation under Grant Nos. AST-0804417, 0334916, 0215793, and 0520822, as well as  
grant NNG05GJ86G from the National Aeronautics and Space  
Administration under the Terrestrial Planet Finder Foundation Science  Program.  
A portion of the research in this paper was carried out at the Jet Propulsion Laboratory, 
California Institute of Technology, under a contract with the National Aeronautics and Space 
Administration and was funded by internal Research and Technology Development funds.
The Lyot Project grateful acknowledges the support of the  
US Air Force and NSF in creating the special Advanced Technologies  
and Instrumentation opportunity that provides access to the AEOS  
telescope.  Eighty percent of the funds for that program are provided  
by the US Air Force.  This work is based on observations made at the  
Maui Space Surveillance System, operated by Detachment 15 of the U.S.  
Air Force Research Laboratory Directed Energy Directorate.
This work has been partially supported by the NSF Science and Technology Center for 
Adaptive Optics, managed by the University of California at Santa Cruz under cooperative 
agreement AST 98-76783.  
The Lyot Project is also grateful to the Cordelia Corporation, Hilary  
and Ethel Lipsitz, the Vincent Astor Fund, Judy Vale and an anonymous  
donor, who initiated the project. 



\bibliography{/Users/shinkley/Desktop/papers/MasterBiblio_Sasha} 
\bibliographystyle{/Users/shinkley/Library/texmf/tex/latex/apj}


\end{document}